# Gold on graphene as a substrate for SERS study

Yingying Wang,[1] Zhenhua Ni,[2] Hailong Hu,[1] Yufeng Hao,[3] Choun Pei Wong,[1] Ting Yu,[1] John TL Thong,[3] and Ze Xiang Shen,[1,a)]

[1]Division of Physics and Applied Physics, School of Physical and Mathematical Sciences, Nanyang Technological University, 21 Nanyang Link, Singapore 637371

[2]Department of Physics, Southeast University, Nanjing 211189, China

[3]Center for Integrated Circuit Failure Analysis and Reliability (CICFAR), Department of Electrical and Computer Engineering, National University of Singapore, Singapore 117576

**Abstract**

In this paper, we report our study on gold (Au) films with different thicknesses deposited on single layer graphene (SLG) as surface enhanced Raman scattering (SERS) substrates for the characterization of rhodamine (R6G) molecules. We find that an Au film with a thickness of ~7 nm deposited on SLG is an ideal substrate for SERS, giving the strongest Raman signals for the molecules and the weakest photoluminescence (PL) background. While Au films effectively enhance both the Raman and PL signals of molecules, SLG effectively quenches the PL signals from the Au film and molecules. The former is due to the electromagnetic mechanism involved while the latter is due to the strong resonance energy transfer from Au to SLG. Hence, the combination of Au films and SLG can be widely used in the characterization of low concentration molecules with relatively weak Raman signals.

Author to whom correspondences should be addressed: zexiang@ntu.edu.sg

Since the first discovery of anomalously strong Raman signals of pyridine adsorbed on rough silver electrodes in 1974,[1] SERS has been widely used to study molecules with low concentrations. In SERS, metals such as Cu, Ag, or Au with rough surfaces are used. Using Au as SERS substrates has advantages because of its chemical stability and excellent biocompatibility.[2,3] To enhance the Raman signals from the adsorbed molecules, Au films with different surface morphologies have been developed. These include electrochemically roughened surface,[4] nanoparticles,[5] plasma-treated island films,[6] as well as nano-structured surfaces.[7] However, the presence of the PL background from Au films[8,9] remains a major obstacle for obtaining molecular vibration information. The PL background can lead to premature saturation of the signal response within the charge-coupled device (CCD) array, thereby limiting acquisition times. Hence, reducing the PL background is important as it will enable longer acquisition times. This improves the signal-to-noise ratio in the Raman spectrum and makes the Raman peaks much more visible. Thus, finding an effective method to quench the PL background from the Au film is important for the SERS technique to be a robust method for detecting trace levels of various chemical compounds.

Graphene, a planar atomic layer of carbon atoms hexagonally arranged, is well known because of its fascinating electronic properties.[10] The use of graphene as quenchers for electronic and vibrational excitations has attracted much attention recently. Graphene has continuous electronic energy levels starting from the Dirac point, which makes it an acceptor for any excitation energy.[11,12] Graphene as a PL quencher can be used for visualizing graphene layers[13] as well as for multicolor fluorescent DNA analysis.[14]

In this work, we observed an efficient quenching of the PL from Au films deposited on SLG. This is due to the resonance energy transfer from Au to SLG. We further combined Au films with SLG to achieve an improved substrate for the enhancement of Raman signals to PL ratio from R6G molecules by suppressing the PL background. We also compared the Raman signals and PL background of R6G adsorbed on Au/SLG substrates, Au substrates, and $SiO_2$/Si substrates. Au/SLG substrates gave the best results.

The graphene samples fabricated by micromechanical cleavage were transferred to $SiO_2$ (300 nm)/Si substrates. The thickness of the graphene layer was determined by white light contrast spectroscopy.[15,16] Au films with different thicknesses were deposited on $SiO_2$/Si substrates containing graphene layers, using a thermal evaporator deposition system (Edwards Auto 306), with a background vacuum level of $5\times10^{-7}$ mbar. The thickness of the film was then monitored with a quartz crystal resonator and checked with an atomic force microscope (AFM). The Raman/PL spectra were acquired by a WITEC CRM200 system using an excitation energy of 532 nm (2.33 eV). The Raman/PL images were obtained by moving the piezo stage with a step size of 100 nm.

Figure 1(a) shows the optical image of SLG on $SiO_2$/Si substrate, while Fig. 1(b) gives the optical image of the SLG after deposition of a 7 nm Au film. Fig. 1(c) presents the PL spectrum of the Au film alone and the PL spectrum of Au on SLG. It can be clearly seen that the PL of the Au film is significantly suppressed by the SLG, ~40.6% of the integrated intensity of the PL is quenched. We also carried out the PL intensity measurements of Au films with other thicknesses, 3 nm, 5 nm, 10 nm and 15 nm. The respective amounts of PL integrated intensities quenched by SLG were 33.2%, 34.2%, 31.9%, 37.2% (results are not shown here). Since the work function of Au, (5.1 eV) is higher than that of graphene (4.46 eV),[17] the charge transfer mechanism cannot be the reason for the PL suppression.[18] Resonance energy transfer from molecules to graphene has been established theoretically and observed experimentally.[11-13] Graphene has continuous electronic levels beyond the Dirac point,[19] which makes it a promising quencher for electronic and vibrational excitations.[11] Therefore, we believe this kind of PL quenching is due to resonance energy transfer from Au to graphene, as schematically shown in Fig.1(c) in the upper right corner. In Fig. 1(c), the vibrational bands from graphene, (G band and 2D band) as well as that from Si (labeled as *) are also observed. Fig. 1(d) gives the imaging of the PL intensity from the Au film. The imaging is obtained by integrating the signal between 195 $cm^{-1}$ and 3890 $cm^{-1}$ ( 537.6 nm to 670.8 nm). Here, a brighter color represents a higher integrated intensity. An obvious contrast can be seen, which shows that there is significant suppression of the PL signal from regions of the Au film in contact with the SLG.

Since SLG can efficiently quench the PL signals from the Au film, this will be important for low concentration molecule detection where Raman signal is weak and difficult detect. We used R6G molecules, which are commonly used in SERS experiments to test the efficiency of this kind of substrates. Au and Au/SLG (bi-layer graphene(BLG)) substrates were soaked in R6G solution (5 µM in water) for 30 minutes. De-ionized water was used to flush away unadsorbed molecules. The substrates were then dried in air. Fig. 2(a) shows the optical image of SLG on $SiO_2$(300 nm)/Si substrate, while Fig. 2(b) gives the optical image of SLG after depositing 7 nm of Au film. The boundary of the SLG is marked by the blue dashed lines. Fig. 2(d) presents the PL image obtained by integrating intensities from 195 $cm^{-1}$ to 3890 $cm^{-1}$. The PL here includes the PL signals from the R6G molecules as well as those from the Au film. A brighter color represents a higher PL/Raman intensity. It can be seen that the PL background is significantly quenched by the SLG. Fig. 2(c) presents the Raman image obtained by summing the intensities of all the vibrational peaks from R6G. These peaks appear at 610.9 $cm^{-1}$, 771.3 $cm^{-1}$, 1122.0 $cm^{-1}$, 1182.2 $cm^{-1}$, 1305.6 $cm^{-1}$, 1356.5 $cm^{-1}$, 1498.4 $cm^{-1}$, 1523.4 $cm^{-1}$, 1573.8 $cm^{-1}$ and 1641.3 $cm^{-1}$. The Raman spectrum of R6G is shown in Fig.3. The detailed assignment of the vibrational peaks from R6G can be found in Ref.12. It can be seen from the Raman image that the Raman intensity of R6G on Au/SLG substrate is slightly weaker than that on Au substrate alone. Attention should be paid to the comparison of Fig. 2(c) with Fig. 2(d). Although the Raman intensity of R6G on Au/SLG substrate is comparable to that on Au substrate, the PL intensity of the former is much weaker than that of the latter. Thus, the combination of Au film and SLG results in a much better substrate than Au film alone for low concentration molecule detection.

Figure 3 gives the Raman spectrum of R6G on Au (7 nm)/SLG substrate, as well as the Raman spectra of R6G on Au film, and on SiO$_2$/Si substrates for comparison. All the spectra given here were collected under the same conditions (laser intensity, integration time, etc). For R6G on SiO$_2$/Si substrate, only PL from the molecules is observed. No Raman signals from the molecules can be detected because of the low concentration of R6G (5 μM in water). On the contrary, for the same concentration of R6G, the Raman signals on Au film are much stronger and detectable. This indicates strong SERS activity with the adsorption of R6G molecules on Au film with the electromagnetic mechanism involved.[20] For molecules adsorbed on Au/SLG substrate, a significant degree of PL quenching can be seen, which is consistent with the results given in Fig. 2(d).

In order to find the substrate with the highest efficiency in obtaining the highest Raman intensities from the molecules by optimizing the Au film as well as the lowest PL background by using SLG to quench PL signal, Raman spectroscopy/imaging studies of R6G on Au film with different thicknesses were carried out and the results are summarized in Figs. 4(a) and 4(b). Each data point presented in these figures is the statistical results of the SERS spectra collected from four different graphene samples. For each graphene sample, three spectra at different positions were taken. For each Au substrate, the results are the statistical results of six spectra from different positions. Fig. 4(a) gives the Raman intensity of R6G which is calculated by summing the intensities of all the Raman peaks belonging to R6G. The Raman intensity of R6G on Au/SLG (BLG) substrate is slightly weaker than that on Au substrate. Fig. 4(b) shows the PL intensity of R6G on Au film obtained by integrating intensities from 195 cm$^{-1}$ to 3890 cm$^{-1}$. The PL background is strongly suppressed by SLG (BLG). We see that the 7 nm Au film is the best candidate for obtaining the highest Raman to PL intensity ratio. Above phenomenon can be understood accordingly. The samples those are less or equal to 7 nm form islands. The 7 nm samples have the highest density and also show very sharp tips, resulting in high SERS signals. Samples 10 nm or thicker form continuous films which have much smoother morphology, hence reduce SERS activity.[21] Based on above observation, it can be seen that the combination of Au film and graphene leads to more sensitive detection because of the efficient suppression of the PL background even though a lightly decrease of SERS intensity.

In summary, Au films with different thicknesses were deposited on graphene layers on SiO$_2$/Si substrate. It is found that SLG can efficiently quench the PL from the Au films. This could be due to the strong resonance energy transfer from Au to graphene. Based on the Raman spectroscopy/imaging measurements, Au/SLG is a much better substrate than Au film alone: the former gives SERS enhancement of the molecules and can efficiently quench the PL background from both the Au. The combination of Au and SLG is a promising SERS substrate which is suitable either for characterization of molecules with low concentration, or molecules with relatively strong PL background.

FIG. 1.

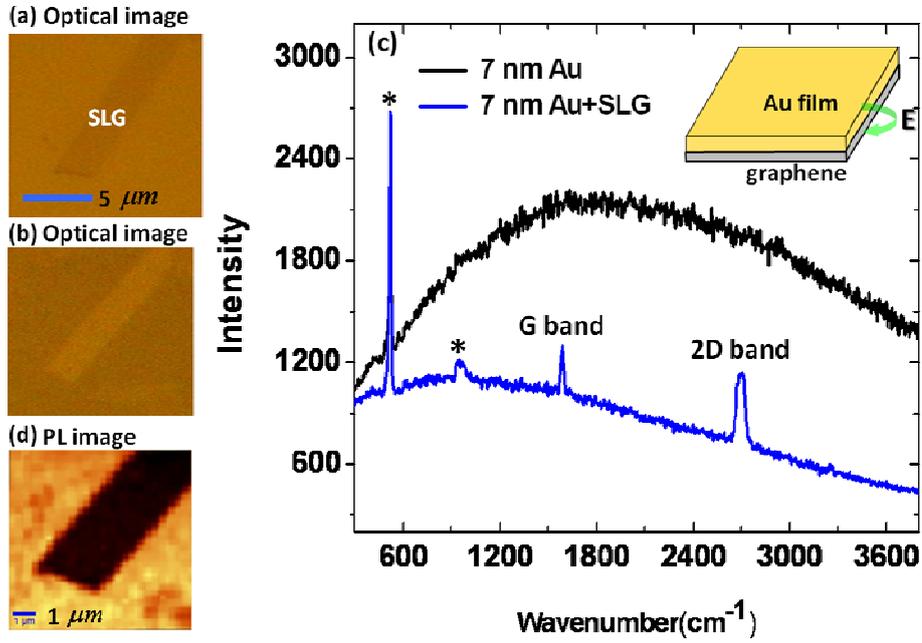

FIG. 1. (a) Optical image of SLG on $SiO_2$ (300 nm)/Si substrate. (b) Optical image of SLG after depositing a 7 nm Au film. (c) PL spectrum from Au film alone (without SLG) and PL spectrum from Au on SLG. (d) PL intensity image. The dark region consists of Au on SLG, and the bright region consists of Au along (without SLG). The image here is obtained by integrating the signals from 195 cm$^{-1}$ to 3890 cm$^{-1}$.

FIG. 2.

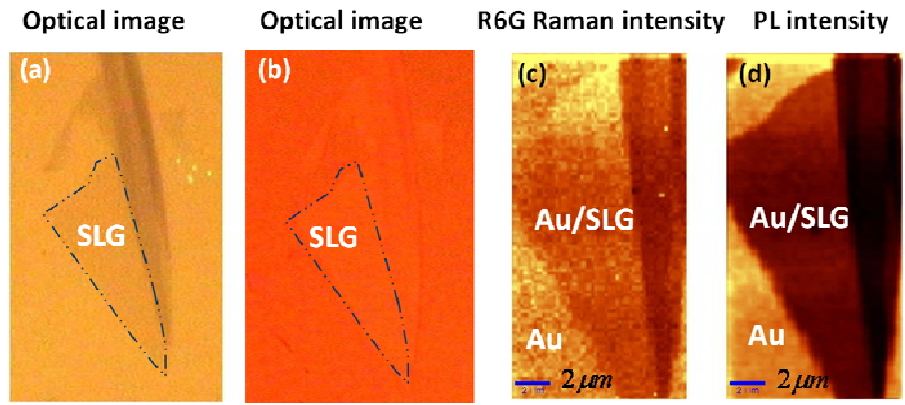

FIG. 2. (a) Optical image of SLG on $SiO_2$ (300 nm)/Si substrate. (b) Optical image of SLG on $SiO_2$/Si substrate after depositing a 7 nm Au film. (c) Raman intensity imaging of R6G. (d) PL intensity image of R6G and Au film. The image is obtained by integrating the signals from 195 $cm^{-1}$ to 3890 $cm^{-1}$.

FIG. 3.

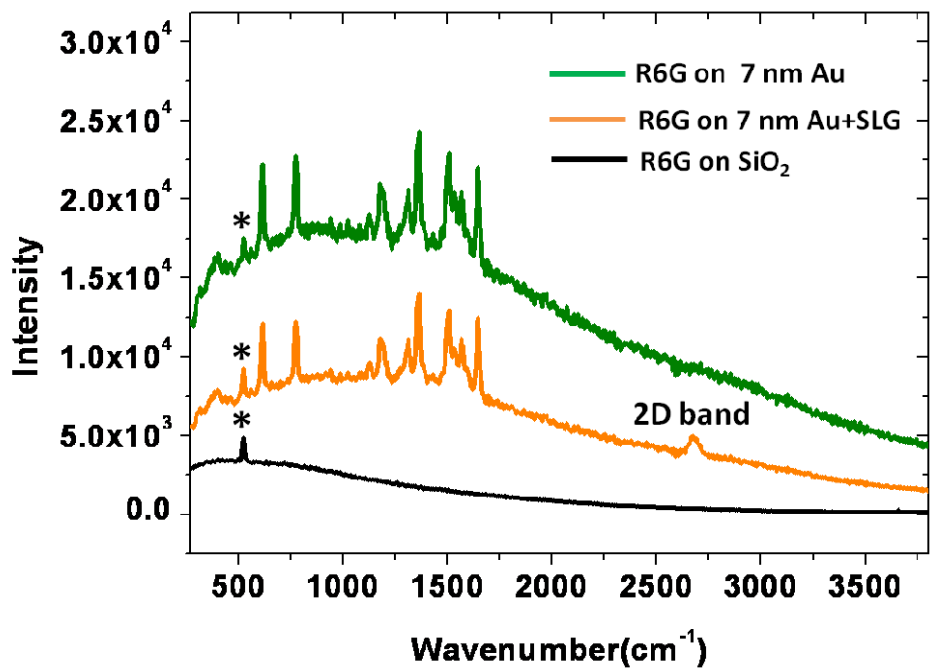

FIG. 3. Raman spectra of R6G on $SiO_2$ substrate, R6G on Au/SLG substrate, as well as R6G on Au substrate.

FIG. 4.

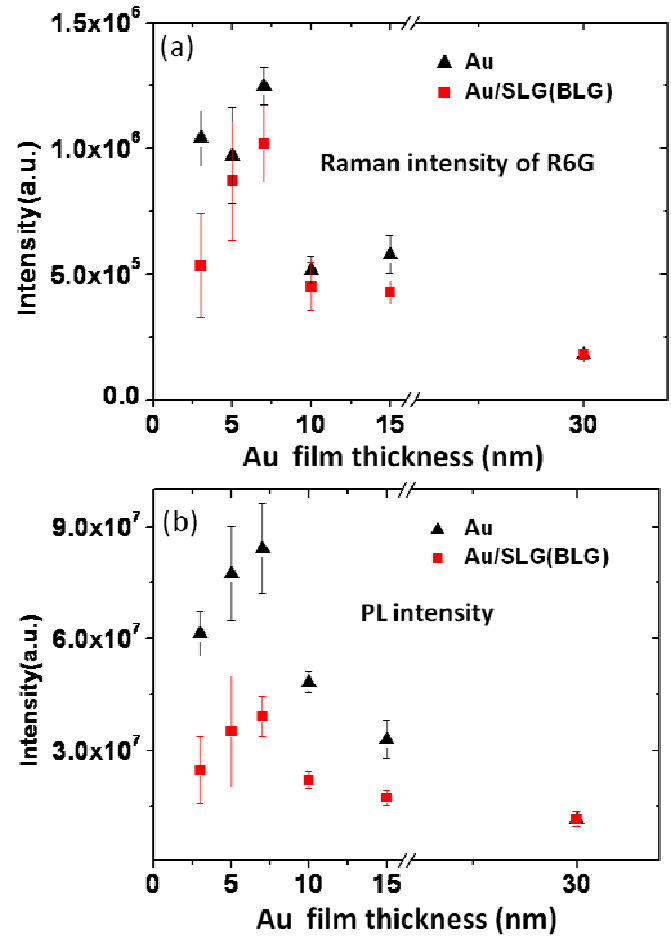

FIG. 4. (a) The Raman intensities of R6G on two substrates: Au substrate and Au/SLG(BLG) substrate. (b) The PL intensities (from R6G and Au film) on two substrates: Au substrate and Au/SLG(BLG) substrate.